\documentclass[aps,prl,twocolumn,groupedaddress,nofootinbib]{revtex4-1}
\usepackage{amsmath}
\usepackage{graphicx}
\usepackage{sistyle}
\usepackage{gensymb}
\usepackage{bm}
\usepackage{natbib}
\usepackage{dsfont}
\usepackage{epsfig}

\usepackage{color}

\begin{document}

\title{Optical levitation of $10$ nanogram spheres with nano-$g$ acceleration sensitivity}

\author{Fernando Monteiro}
\email{fernando.monteiro@yale.edu}
\author{Sumita Ghosh} 
\author{Adam Getzels Fine}
\author{David C. Moore}
\affiliation{Wright Laboratory, Department of Physics, Yale University, New Haven, CT 06520, USA}

\begin{abstract}
We demonstrate optical levitation of SiO$_2$ spheres with masses ranging from 0.1 to 30 nanograms. In high vacuum, we observe that the measured acceleration sensitivity improves for larger masses and obtain a sensitivity of $0.4 \times 10^{-6}\ g/\sqrt{\mathrm{Hz}}$ for a 12 ng sphere, more than an order of magnitude better than previously reported for optically levitated masses.  In addition, these techniques permit long integration times and a mean acceleration of $(-0.7\pm2.4\,[stat] \pm 0.2\,[syst])\times ~ 10^{-9}\,g$ is measured in $1.4\times 10^4$~s. Spheres larger than 10~ng are found to lose mass in high vacuum where heating due to absorption of the trapping laser dominates radiative cooling.  This absorption constrains the maximum size of spheres that can be levitated and allows a measurement of the absorption of the trapping light for the commercially available spheres tested here. Spheres consisting of material with lower absorption may allow larger objects to be optically levitated in high vacuum.  
\end{abstract}

\maketitle

\section{Introduction}

Optically trapped dielectric spheres~\cite{Ashkin:1971,Ashkin:1986} have enabled
a wide variety of precision sensing applications ranging from biology (e.g.,~\cite{Ashkin:1987_science,Ashkin:1987_nature,Neuman:2004}) to fundamental physics~\cite{Geraci:2010,Rider:2016,Moore:2014}. For objects levitated in high 
vacuum, excellent acceleration sensitivity is possible since they 
can be isolated from environmental sources of thermal noise, eliminating the
primary source of dissipation present for most force sensors and accelerometers.
Optically levitated microspheres and nanospheres are currently being investigated for 
applications in optomechanics~\cite{Li:2011,Millen:2015,Vovrosh:2017}, tests of the quantum mechanical properties of massive objects~\cite{Romero:2010,Kaltenbaek:2012}, precision force sensing~\cite{Yin:2013,Ether:2015,Ranjit:2015,Ranjit:2016,Jain:2016,Hempston:2017,Rider:2017}, and searches for new fundamental interactions~\cite{Geraci:2010,Moore:2014,Rider:2016}.

Applications of levitated microspheres that search for forces that couple to mass~\cite{Geraci:2010,Rider:2016} or the number of atoms or nucleons in the sphere~\cite{Moore:2014} require optimizing the sensitivity to accelerations acting on the test mass. 
Although other precision accelerometers employing macroscopic masses~\cite{Kapner:2007} or atom interferometry~\cite{Dickerson:2013} can reach smaller acceleration sensitivities, the techniques described here are unique in the microscopic scale of the accelerometer and the ability to thermally and electrically isolate the test mass from the environment.  Thus, optically levitated masses provide a powerful technology for searching for new interactions producing accelerations at distances $\lesssim 100\ \mu$m.

Force sensitivity as low as a few zeptonewtons has been previously demonstrated for optically levitated objects~\cite{Ranjit:2016,Jain:2016,Hempston:2017,Vovrosh:2017}.  However, the small size of the masses used in these systems (typically 50-500~nm) leads to acceleration sensitivities of $10^3$\textendash$10^6$~$\mu g/\sqrt{\mathrm{Hz}}$, where $g = 9.8$~m/s$^2$ is the acceleration due to gravity.  For levitated microspheres larger than 1~$\mu$m, previous work reached sensitivity of 7.7~$\mu g/\sqrt{\mathrm{Hz}}$ for $d=5$~$\mu$m spheres using a single-beam heterodyne detection scheme~\cite{Rider:2017}, and 47~$\mu g/\sqrt{\mathrm{Hz}}$ for $d=3$~$\mu$m spheres in a multi-beam feedback system~\cite{Li:2011}.
Here we present the smallest acceleration sensitivity reported to-date for an optically levitated particle by more than an order of magnitude. Lower sensitivities are achieved through the use of an optical trap that can levitate large ($\gtrsim 20\ \mu$m diameter) SiO$_2$ spheres in high vacuum.  In addition, a 
weakly converging trapping laser is used to provide a small optical spring constant.
The use of massive spheres provides a large enhancement in the signal-to-noise of the
optical detection system due to the large amount of scattered light, while the weak trap allows a larger center of mass displacement for a given applied force. For the largest spheres tested, an acceleration sensitivity at the nano-$g$ scale is reached in a measurement time of \SI{E4}{s}.

\section{Experimental Setup}

\begin{figure}[t]
\centering
\includegraphics[width=\columnwidth]{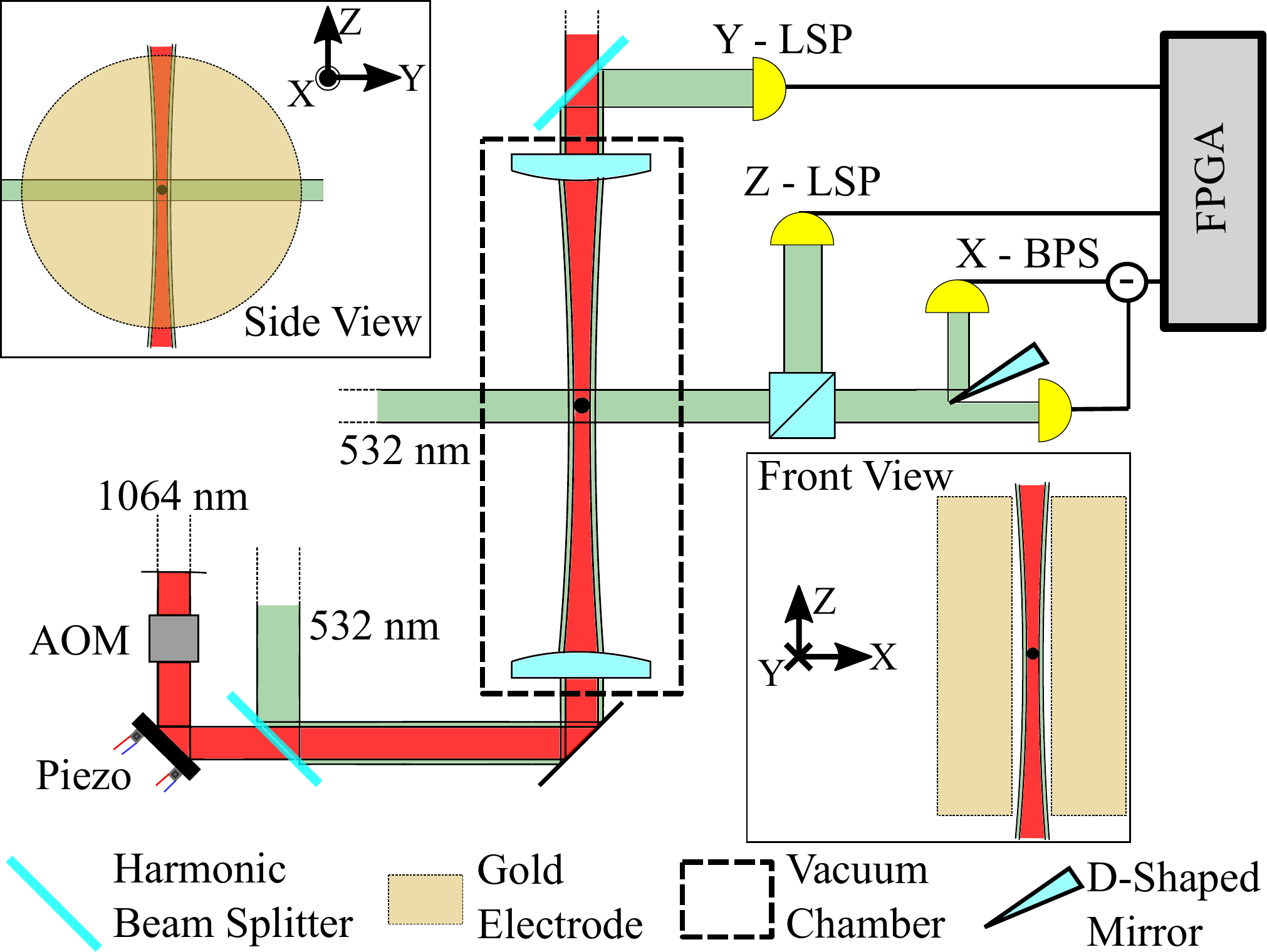}
\caption{Simplified schematic of the experimental setup.  The microsphere is levitated by a \SI{1064}{nm} laser (dark red) and imaged by two \SI{532}{nm} beams (light green). Signals coming from the imaging sensors are fed into an FPGA that provides active feedback using the AOM and piezo deflection mirror. The insets show the front and side view of the levitated sphere with respect to the electrodes used to generate the electric field for calibration. The vertical direction is defined as Z, while X denotes the direction perpendicular to the electrode's surface. The electrode diameters are \SI{25.4}{mm} and the electrode separation used in these measurements is $\sim$\SI{2}{mm}.}
\label{schema}
\end{figure}

A simplified schematic of the experimental setup is shown in Fig.~\ref{schema}.  Microspheres are levitated using a weakly focused vertical trapping beam with wavelength $\lambda$\,=\,\SI{1064}{nm} and numerical aperture NA\,=\,$0.03$. Following \cite{Ashkin:1971,Moore:2014} a dielectric sphere can be levitated at a stable equilibrium position above the focus of the beam. The equilibrium height for a given microsphere can be adjusted by varying the laser power to balance its weight as the beam diverges above the focus. The trap is operated inside a vacuum chamber at pressure down to $\sim~10^{-7}$~mbar in order to minimize the center of mass (CM) motion caused by collisions with residual gas molecules.

Two additional beams with $\lambda$\,=\,\SI{532}{nm} are used to image the motion of the sphere in the vertical and radial directions. These imaging beams have a waist that is much larger than the sphere size in order to fully illuminate the sphere. To monitor the size and position of the spheres, aspheric lenses are focused on the trap location in the vertical and horizontal planes, and the light transmitted past the microsphere is used to form orthogonal microscope images at the output of the vacuum chamber. For the X direction, as defined in Fig.~\ref{schema}, this image is split in half by a D-shaped pickoff mirror and measured using a balanced photodiode (BPD) to minimize readout noise~\cite{Li:2011}. The Y and Z directions are imaged by a lateral effect position sensor with larger dynamic range.

The outputs of these sensors are fed into an active feedback loop that is used to damp the microsphere's motion as the pressure is reduced~\cite{Ashkin:1977,Li:2011,Moore:2014}.  The imaging signals are fed into a field-programmable gate array (FPGA), which provides the control signals for the feedback loop.  The feedback is provided for the Z degree of freedom by modulating the trapping beam power using an acousto-optic modulator (AOM).  Feedback in the X and Y degrees of freedom is applied by displacing the trapping beam using a high-bandwidth piezo deflection mirror at frequencies up to $\sim 1$~kHz. Feedback is required to maintain stable trapping of the microsphere for pressures $\lesssim 0.1$~mbar, where damping from the residual gas is insufficient to prevent heating of the CM motion of the sphere by the laser.

For this detection system, the best acceleration sensitivity is obtained for a small optical restoring force, which allows a larger CM displacement for a given excitation. The restoring force is tuned using the laser power to adjust the equilibrium position of the sphere above the focus.  For the results reported here, a beam waist of \SI{25}{\mu m} was measured at the sphere's equilibrium position, located 730~$\mu$m above the focal point of the trapping beam. The large beam waist allows microspheres with diameters ranging from 5\textendash30~$\mu$m to be trapped at the same equilibrium height and provides resonant frequencies for the spheres' motion in the trap between 100 to \SI{200}{Hz}, depending on the sphere size.  

\begin{table}
\begin{ruledtabular}
\caption{Properties of the SiO$_2$ microspheres used in this work. The distribution of sphere diameters provided by the microsphere manufacturer, $d_{sup}$, are compared to the measurement of the diameter distribution performed here, $d_{meas}$, where the table reports the mean and standard deviation of the sphere size distribution.  The density, $\rho$, assumes the value provided by the manufacturer.}
\label{table1}
\begin{tabular}{ c  c  c  c }
   $d_{sup}\,$[$\mu$m] & $d_{meas}\,$[$\mu$m] & $\rho\,$[g$/$cm$^3$] & Manufacturer \\ \hline
   5.0 $\pm$ 0.2 & 5.0 $\pm$ 0.2 & 2.0 & Bangs Laboratories\footnote{http://www.bangslabs.com} \\
   10 & 10.3 $\pm$ 1.4 & 1.8 & Corpuscular\footnote{http://www.microspheres-nanospheres.com} \\
   15 & 15.0 $\pm$ 2.7 & 1.8 & Corpuscular \\
   22.62 $\pm$ 0.76 & 22.7 $\pm$ 0.7 & 1.85 & Microparticles GmbH\footnote{http://microparticles.de} \\
   32 & 30.9 $\pm$ 3.1 & 1.8 & Corpuscular
\end{tabular}
\end{ruledtabular}
\end{table}

The silica microspheres used in this work have mean diameter ranging from 5 to \SI{32}{\mu m} and are supplied by different manufacturers. Table~\ref{table1} lists the diameters reported by each manufacturer, which are compared to the diameter distribution measured in this work.  To validate the manufacturer diameter specification, a calibrated optical microscope image containing $\gtrsim$\,$10^2$ microspheres was analyzed to measure the sphere size distribution.  Our independent measurements of the mean and width of this distribution are in good agreement with the manufacturer specifications, where available.  The density of the silica spheres reported by each manufacturer (varying from 1.8\textendash2.0~g/cm$^3$) is also listed in Table~\ref{table1} and assumed throughout this work.

\section{Acceleration measurements}

\begin{figure}[t]
\centering
\includegraphics[width=\columnwidth]{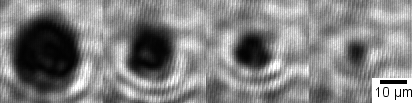}
\caption{Microscope images of microspheres with several different diameters levitated in the optical trap. From left to right the sphere diameters are measured to be $23.0\pm 1.1\,\mu$m, $14.4\pm 0.8\,\mu$m, $11.1\pm 0.7\,\mu$m and $5.0\pm 0.5\,\mu$m. All images have the same scale, indicated on the bottom right, and were obtained with the spheres levitated at the
same equilibrium position above the focus.}
\label{size}
\end{figure}

To determine the mass of a specific microsphere, the size for each sphere is characterized from its measured diameter while trapped using the microscope image at the output of the vacuum chamber, as shown in Fig.~\ref{size}. To calibrate these images, several spheres from the sample containing the $22.62\pm 0.76\,\mu$m microspheres were trapped and used to translate the pixel count observed at the imaging camera to the diameter in $\mu$m.  The $d = 23\ \mu$m diameter microspheres were used as a calibration since they have the minimum relative variance in their diameter distribution. This residual diameter variance and the blurriness of the sphere edges provide the dominant systematic errors from the images, corresponding to an uncertainty on the diameter ranging from 5\% for the $d = 23\ \mu$m spheres up to 10\% for the $d = 5\ \mu$m spheres. This calibration is used to determine the diameter of all spheres in the range from 5 to \SI{32}{\mu m}.

Once a microsphere is trapped at low pressure, the data from the BPD are recorded to determine the microsphere position versus time, which is converted into force or acceleration 
by applying a known electric force.
The calibration is performed by first discharging each sphere~\cite{Moore:2014,Frimmer:2017} until it has a net charge of a single electron using an ultra-violet (UV) lamp, which ejects electrons from the sphere or nearby surfaces. The microsphere's charge can be controllably varied in either polarity.  To remove electrons from the sphere, the UV light is focused on the sphere, while all other surfaces are removed from the vicinity of the trap.  Electrons can be added to the sphere by illuminating a gold electrode with the UV light, which can be positioned near the trapping region using an in-vacuum stage.  The sphere's charge is monitored by measuring the response of the CM motion in the X direction and in the presence of an oscillating electric field that is generated by two parallel electrodes centered around the trapping region, as shown in Fig.~\ref{schema}. 

The voltage amplitude spectral density, $\sqrt{S_V}$ of the BPD output signal in units of V/$\sqrt{\mathrm{Hz}}$ is converted to an acceleration spectral density (ASD), $\sqrt{S_a}$, using the applied electric field, $E$, and the measured charge and mass of the sphere. The electrode spacing was determined by direct measurement to be $2.1\pm0.1$~mm, giving a typical amplitude of $E$=10\textendash30~V/mm during calibration.

\begin{figure}[t]
\centering
\includegraphics[width=\columnwidth]{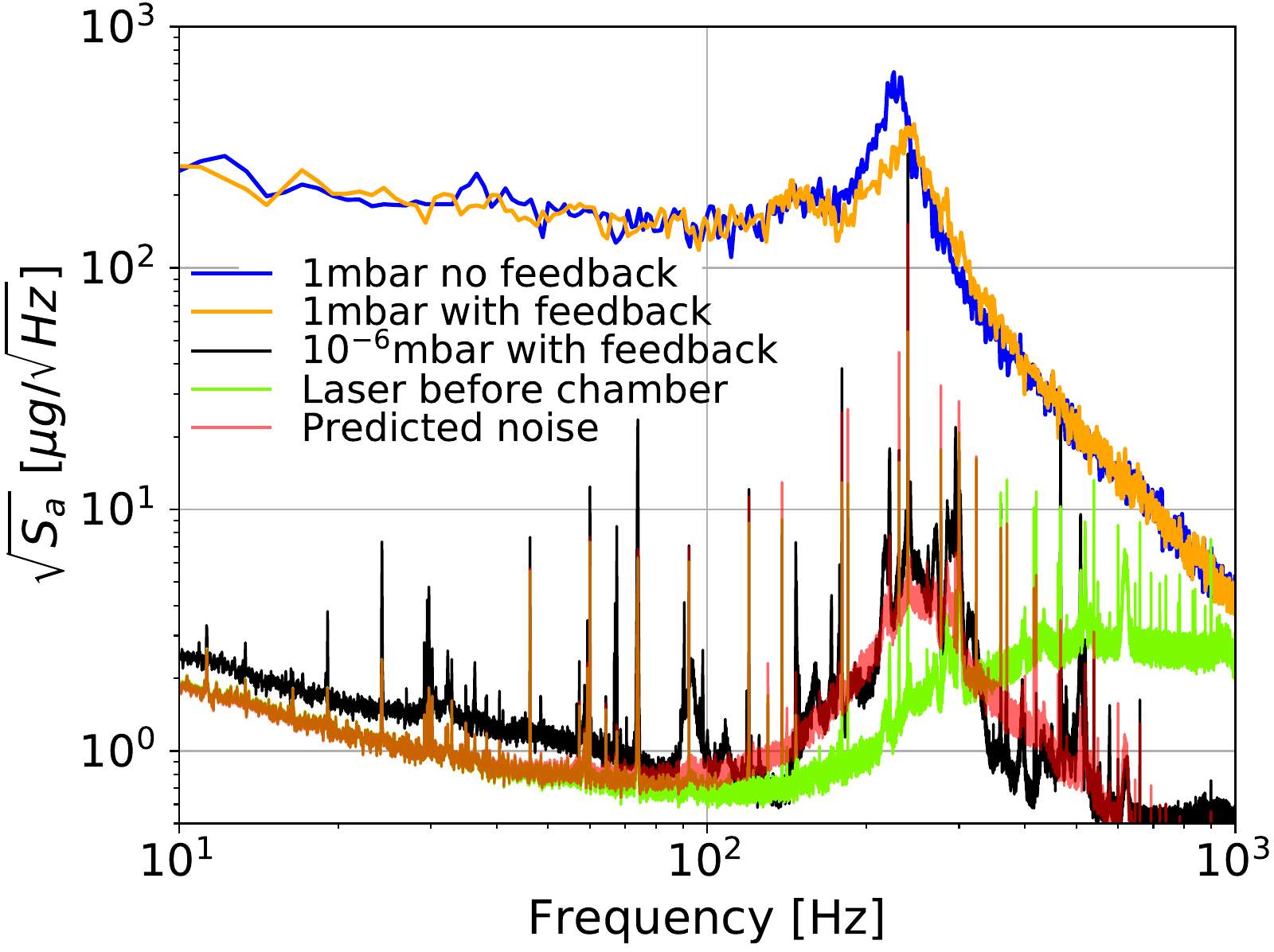}
\caption{ASD measured for a sphere with diameter \SI{14.0}{}$\pm$\SI{0.8}{\mu m} at high and low pressure. The blue line (top) shows the ASD measured at \SI{1}{mbar} with no feedback in the X direction. The orange line (top) shows the damping of the motion near the resonance peak when the feedback in the X direction is turned on, which is necessary to keep the sphere stably trapped at pressures $\lesssim 0.1$~mbar. The black line (bottom) shows the measured ASD at $\approx\,$\SI{E-6}{mbar} where the corresponding acceleration sensitivity is $\lesssim$\,\SI{2}{\mu \textit{g}/\sqrt{Hz}} in the frequency range between \SI{10}{} to \SI{200}{Hz}. The light green curve (bottom) shows the measured spectral density of the pointing fluctuations of the \SI{1064}{nm} laser before it enters the vacuum chamber. These pointing-induced fluctuations are converted to an expected acceleration using the measured transfer function of the sphere (red bottom).}
\label{psd}
\end{figure}

After calibrating the response for each microsphere, the sphere is discharged to have no net charge and the oscillating electric field is turned off. Figure~\ref{psd} shows the ASD measured in the X direction for a sphere with a mass of \SI{2.7}{ng}, which corresponds to a diameter of \SI{14}{\mu m}. The top blue and top orange curves show the ASD measured at pressure $p = 1$~mbar before and after applying the X feedback at frequencies around the resonant frequency. The bottom black curve shows the ASD measured at $\approx\,$\SI{E-6}{mbar} with the X feedback applied. The measured ASD for this sphere is $\lesssim$\SI{2}{\mu \textit{g}/\sqrt{Hz}} in the frequency range between 10 to \SI{200}{Hz}, reaching a minimum below \SI{1}{\mu \textit{g}/\sqrt{Hz}} near 100~Hz. 

Figure~\ref{psd} also shows a measurement of the spectral density of the angular pointing of the trapping beam (bottom light green) before it enters the vacuum chamber, which is measured simultaneously with the bottom black curve. 
The effective pointing-induced motion is calibrated using measurements of the microsphere's CM acceleration induced by an applied oscillatory displacement of the beam pointing using the piezo deflection mirror. For a microsphere trapped at low pressure, the amplitude of the pointing spectrum increases above \SI{200}{Hz} as the gain in the feedback loop increases. The feedback system is tuned to provide negligible response in the 10\textendash100~Hz frequency range where typical measurements are performed, which was verified by observing no change in the beam pointing at these frequencies when the feedback gain was reduced by more than an order of magnitude. The bottom red curve shows the expected contribution of the beam pointing fluctuations before the chamber to the sphere's ASD. This curve is obtained by multiplying the light green curve on the bottom by the sphere's transfer function obtained at low pressure.  The transfer function was directly measured using the response of the sphere's motion to an applied electric field that produced a flat (white noise) amplitude spectrum at frequencies up to the 2~kHz bandwidth of the high voltage amplifier. 
As shown in Fig.~\ref{psd}, the measured CM ASD in the 10\textendash100~Hz range is consistent with the expected acceleration due to the measured pointing noise of the trapping beam before the chamber. The relative beam position stability, defined as $\Delta\alpha = 2\sigma_{x}/w_{0}$~\cite{Kwee:2007}, where $\sigma_{x}$ is the standard deviation of beam position and $w_{0}$ is the beam waist before the vacuum chamber, is calculated to be $\Delta\alpha\approx$~\SI{2e{-6}}{/\sqrt{Hz}} at 50~Hz. It is expected that the pointing noise of this system could be substantially improved with additional passive or active beam stabilization~\cite{Kwee:2007}. 

\section{Results}

\begin{figure}[t]
\centering
\includegraphics[scale = 0.5]{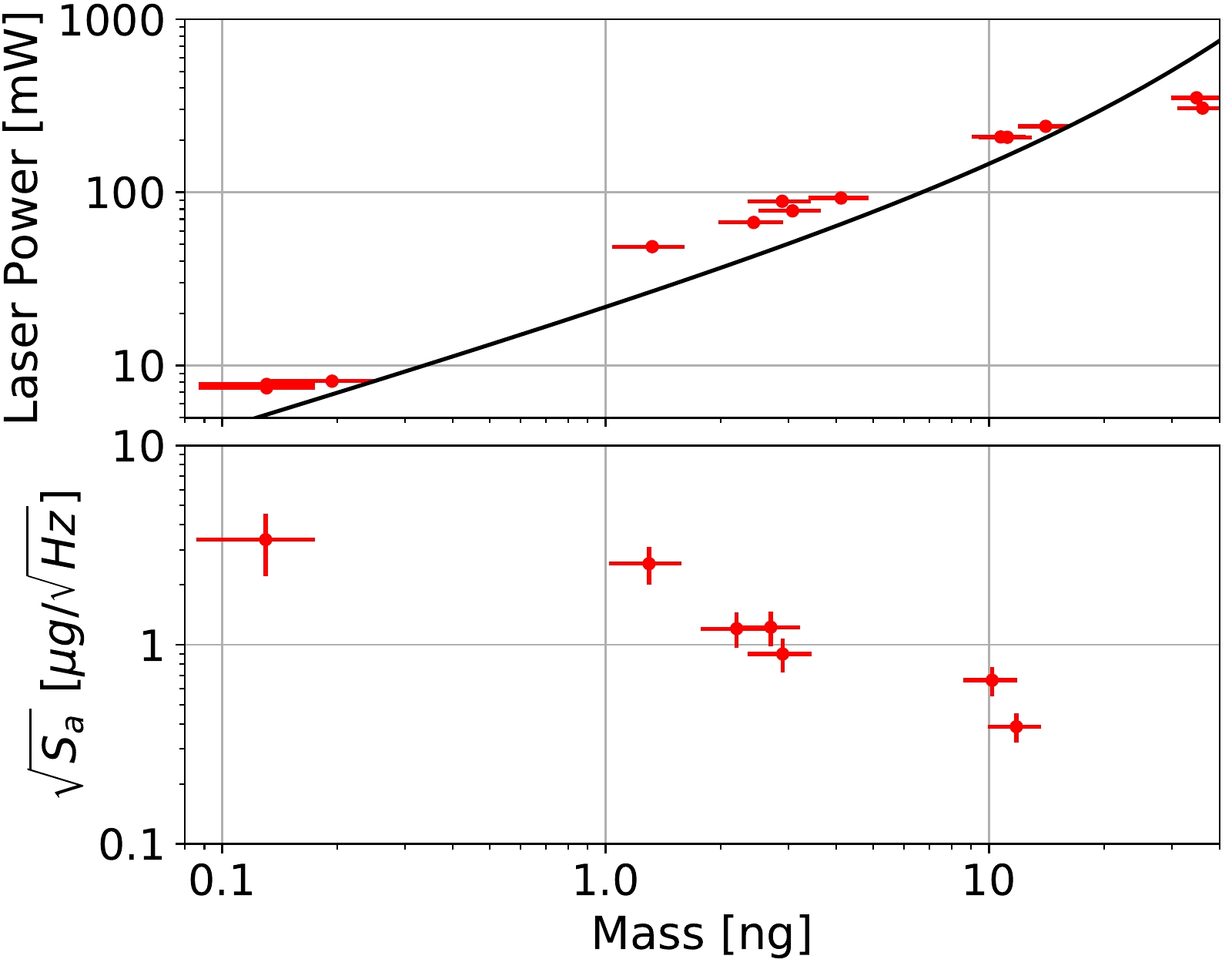}
\caption{(top) Required laser power entering the vacuum chamber to levitate SiO$_2$ spheres of varying masses at \SI{1}{mbar}. The black curve indicates the predicted power in the ray-optics limit ($d \gg \lambda$) using the measured beam parameters. (bottom) Measured ASD for different sphere sizes at a frequency of 50~Hz.}
\label{sensitivity}
\end{figure}

Figure~\ref{sensitivity} (top) shows the laser power required to levitate microspheres with a
variety of masses at a fixed equilibrium position. All powers are compared at a consistent pressure, $p = 1$~mbar, since not all spheres were pumped to lower pressure. The measured power may differ slightly at lower pressures due to secondary effects such as the interaction between the residual gas molecules and the heated sphere surface~\cite{Millen:2014}. The trapping beam power varies from 7~mW for $d = 5\ \mu$m spheres to $> 200$~mW for $d = 30\ \mu$m spheres. The black line shows a calculation of the expected power required using the measured beam waist
and equilibrium position.  This calculation has no free parameters and numerically integrates the scattering of light from the sphere in the ray-optics limit ($d \gg \lambda$)~\cite{ASHKIN:1992}. Reasonable agreement is found between the predicted and measured power over most of the size range considered here, although this simple model somewhat underestimates the required power for spheres $\lesssim 10$~ng, and overestimates the power for the largest spheres.  Mie scattering effects or photothermal forces arising from the residual gas present at 1~mbar~\cite{Ashkin:1976,Millen:2014,Ranjit:2015} may account for the residual discrepancies between the data and model.  Uncertainties in the measurements of the beam waist, numerical aperture, and refractive index of the spheres were not found to provide significant uncertainty in the predicted power.

Figure~\ref{sensitivity} (bottom) shows the measured ASD at a frequency of 50~Hz for a smaller number of spheres that were trapped at low pressure and electrically discharged following the procedure described above. The measured ASDs for all spheres were found to have a similar spectral shape to that shown for the sphere in Fig.~\ref{psd}, with the minimum sensitivity occurring at frequencies ranging from \SI{10}{Hz} to \SI{200}{Hz}.
The inferred acceleration sensitivity is empirically found to improve as the mass increases from 3.7~$\mu g/\sqrt{\mathrm{Hz}}$ for the smallest sphere trapped at low pressure (with $d=5\ \mu$m) to 0.4~$\mu g/\sqrt{\mathrm{Hz}}$ for the largest sphere (with $d=23\ \mu$m). The reported accelerations are one order of magnitude above the sensor and electronics noise contribution.

\begin{figure}[t]
\centering
\includegraphics[scale = 0.5]{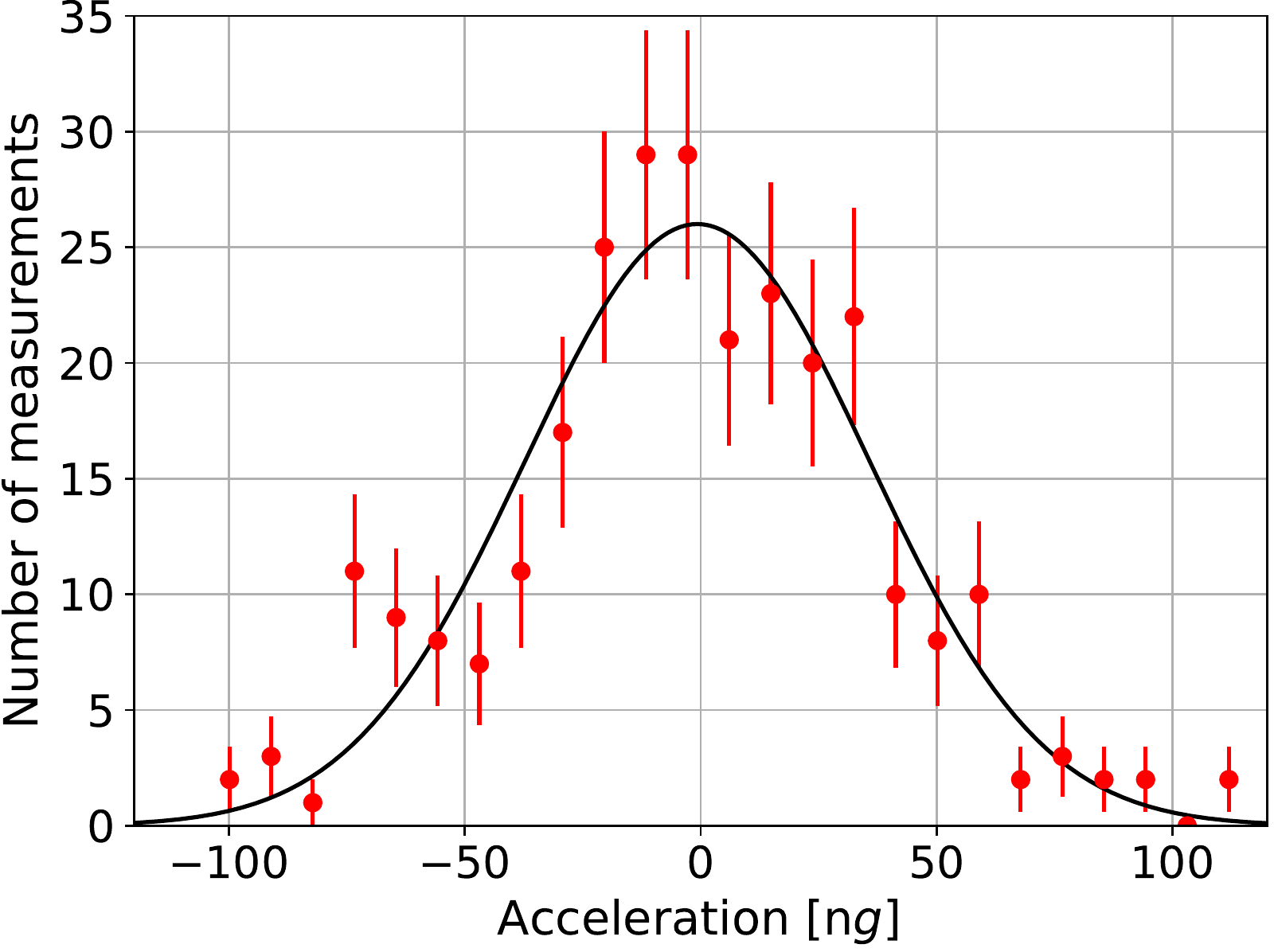}
\caption{Distribution of the acceleration measured for an \SI{12}{ng} sphere for data acquired in individual \SI{50}{s} periods over a total integration time of \SI{1.4E4}{s}. The black line shows a Gaussian fit to the data.}
\label{acc_corr}
\end{figure}

The trap described here can allow stable levitation of a single microsphere for month long time scales, enabling long integrations to reach $\lesssim$\,n$g$ sensitivities.
The acceleration sensitivity for a discharged microsphere was also measured in the absence of an externally applied acceleration by correlating the measured sphere motion with an empirical template for an expected excitation force. The empirical template was determined by the application of a known oscillating electric field to a charged microsphere at a frequency of 50~Hz. Figure~\ref{acc_corr} shows the distribution of the measured accelerations for repeated individual \SI{50}{s} long integrations for an \SI{12}{ng} sphere.  A Gaussian fit to the width of this distribution agrees with the ASD shown in Fig.~\ref{psd} and the mean acceleration is measured to be $(-0.7\pm2.4\,[stat] \pm 0.2\,[syst])\times 10^{-9}\,g$ in a total integration time of \SI{1.4E4}{s}.
While longer integration times are possible, these results already demonstrate the smallest directly measured acceleration for an optically levitated mass~\cite{Ranjit:2015,Ranjit:2016,Rider:2017}.
The dominant systematic errors for this measurement are due to the uncertainty on the distance between the electrodes and the error on the measured diameter from which the sphere mass is calculated.  For the sphere mass calculation, the nominal density measured by the manufacturer is used.  An additional error is conservatively included to account for a possible change in the mass or density of the spheres due to heating by the laser. The laser power required to levitate the spheres at low pressure, where laser heating can be significant, changes by less than 10\% when compared to its value at \SI{1}{mbar}. While such power variation can be due to changes in optical properties and photothermal forces, we conservatively assume the change in levitation power is entirely due to a variation in mass, and include this as an additional systematic error on the mass of the sphere.

\section{Levitation of larger spheres}
\begin{figure}[t]
\centering
\includegraphics[scale = 0.7]{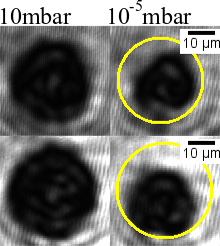}
\caption{Two spheres with initial diameters of \SI{28}{\mu m} and \SI{33}{\mu m} respectively (left column from top to bottom) are observed to decrease in size as the pressure is reduced from \SI{10}{mbar} (left column) to \SI{E-5}{mbar} (right column). The final diameters are \SI{22}{\mu m} and \SI{24}{\mu m} respectively (right column from top to bottom). The yellow circles in the low pressure images represent the corresponding size of spheres at \SI{10}{mbar}.}
\label{32um}
\end{figure}

Following the demonstrated improvement in acceleration sensitivity with mass shown in Fig.~\ref{sensitivity}, microspheres with larger mean diameter of \SI{31}{\mu m} were trapped, corresponding to a mass of $\approx\,$\SI{29}{ng}. The laser power used to trap these spheres is $\sim 330~$mW, corresponding to an intensity of $\sim 0.17~$mW/$\mu$m$^2$ at the sphere location. As shown in Fig.~\ref{32um}, these spheres become smaller as the pressure is quickly reduced from $\sim$\SI{0.1}{mbar} to \SI{E-5}{mbar} using a turbopump.  The final diameter measured after reducing the pressure is \SI{\approx 22}{\mu m}.

The observed reduction in the size of the microspheres apparent in the microscope images corresponds to a simultaneous factor of $\sim$2 reduction in the optical power required to maintain the sphere at the same equilibrium position. Additionally, while the reduction in size was repeatable for two different spheres, both spheres were stably trapped only for several minutes at $p \lesssim$\SI{E-5}{mbar} and their acceleration sensitivity could not be characterized in detail. This behavior puts a practical upper bound on the maximum size of SiO$_2$ spheres that can be optically levitated using these techniques in high vacuum. Such a size reduction is only observed for the sample containing spheres with diameter of 31 um, while the diameter for all smaller spheres is constant within the error of our measurement.

The observed reduction in size is consistent with vaporization of the SiO$_2$ spheres as the pressure is reduced and cooling from the residual gas becomes ineffective. 
In the low pressure limit, the cooling rate of the sphere is assumed to be 
dominated by blackbody radiation, and its equilibrium size is determined by the
expected vaporization temperature for silica at $p = 10^{-5}$~mbar~\cite{Lamoreaux:1987}, $T_{vap} \approx 1360$~K.  Assuming a total emissivity for silica at this temperature of $\epsilon \approx 0.4$~\cite{Sully:1952,Sova:1992} and setting the absorbed power equal
to the radiated power at $T_{vap}$ provides an estimated heating due to
absorption of the trapping laser light of $\approx 0.1$~mW. This heating
corresponds to an optical absorption coefficient of $\approx$\,\SI{3E-5}{\mu m^{-1}} at $\lambda = 1064$~nm.

The above calculation considers the levitated particle as a surface emitter due to its large diameter of \SI{\sim 31}{\mu m} compared to the peak wavelength of the emitted radiation, which is \SI{\sim 2}{\mu m} at $T_{vap}$~\cite{Schweiger1994,Takehiro2009}. Spheres with radius comparable to the thermal radiation wavelength will act as a volume emitter, leading to a higher cooling rate relative to surface emission alone~\cite{Schweiger1994}.

Future characterization of the sphere's temperature and absorption will provide a better understanding of this size reduction, and if confirmed, suggests spheres with lower optical absorption may allow optical levitation of larger masses in high vacuum.  The inferred absorption coefficient for the 30~$\mu$m diameter spheres is more than 2 orders of magnitude larger than for optical grade fused silica, possibly due to the inclusion of water or other impurities in the sphere. In particular, production of commercial monodisperse silica microspheres such as those used in this work typically follows the St\"{o}ber process~\cite{STOBER:1968}, for which substantial content of water and silanol groups~\cite{Zhang:2009} leads to the $\sim$10\textendash20\% lower densities given in Table~\ref{table1} relative to the density of optical grade fused silica.  
Future work will investigate the use of microspheres produced with methods that do not introduce substantial water impurities, and could enable more massive objects to be levitated
in high vacuum.

\section{Conclusion}
We have demonstrated the optical levitation of SiO$_2$ microspheres with masses ranging from 0.1 to 30 nanograms (corresponding to diameters between 5 and \SI{33}{\mu m}). The measured acceleration sensitivity was found to improve for spheres with larger masses and a mean acceleration of $(-0.7\pm2.4\,[stat] \pm 0.2\,[syst])\times 10^{-9}\,g$ was measured for a 12 ng sphere in an integration time of \SI{1.4E4}{s} in the absence of any externally applied forces. The corresponding acceleration sensitivity is the best reported to-date for an optically levitated object.

The acceleration sensitivity for the current apparatus was determined to be limited by pointing fluctuations of the beam used to levitate the spheres. Future work to stabilize the trapping laser could allow substantially smaller acceleration sensitivities to be reached.  Such high sensitivity combined with the microscopic scale of the accelerometers described here can enable high-precision searches for new fundamental interactions at short distance, including searches for new short-range forces that couple to mass~\cite{Geraci:2010,Rider:2016} or electric charge~\cite{Moore:2014}.

Silica microspheres with diameters larger than 20~$\mu$m were found to decrease in
mass in high vacuum, consistent with vaporization from heating due to absorption of the trapping laser light at high vacuum pressures where cooling from the residual gas is negligible. The inferred absorption coefficient for the microspheres tested here is substantially larger than optical grade fused silica, likely due to the presence of water and other impurities within the spheres.  Optical levitation of larger objects may be possible for microsphere materials with lower optical absorption.

\section{Acknowledgments}
We would like to thank C.~Blakemore, G.~Gratta, A.~Kawasaki, and A.~Rider (Stanford) for useful discussions related to this work. This work is supported, in part, by the Heising-Simons Foundation, NSF Award No. 1653232, and Yale University.

\bibliographystyle{apsrev4-1}
\bibliography{refe}{}

\end{document}